\documentclass[preprint,tightenlines,showpacs,nofootinbib,aps,prc,amsmath,amssymb,floatfix]{revtex4-1} 
\usepackage{graphicx}% Include figure files
\usepackage{bm}% bold math
\usepackage{url,multirow,xcolor}
\begin{document}

\title{Reproduction of exact solutions of Lipkin model by nonlinear higher random-phase approximation}

\author{ J.~Terasaki,$^1$ A.~Smetana,$^1$ F.~\v{S}imkovic,$^{1,2,3}$  and M.~I.~Krivoruchenko$^{3,4,5}$ }
\affiliation{ $^1$Institute of Experimental and Applied Physics, Czech Technical University in Prague, Horsk\'{a} 3a/22, 128 00 Prague 2, Czech Republic }
\affiliation{ $^2$Department of Nuclear Physics and Biophysics, Comenius University, Mlynsk\'{a} dolina F1, SK-842 48 Bratislava, Slovakia}
\affiliation{ $^3$Bogoliubov Laboratory of Theoretical Physics, Joint Institute for Nuclear Research, 141980 Dubna, Moscow Region, Russia}
\affiliation{$^4$Institute for Theoretical and Experimental Physics, B. Cheremushkinskaya 25, 117218 Moscow, Russia}
\affiliation{$^5$Department of Nano-, Bio-, Information, and Cognitive Technologies, Moscow Institute of Physics and Technology, 9 Institutskii per. 141700 Dolgoprudny, Moscow Region Russia}

\date{Oct.~5, 2017}
\begin{abstract} 
It is shown that the random-phase approximation (RPA) method with its nonlinear higher generalization, which was previously considered as approximation except for a very limited case,  reproduces the exact solutions of the Lipkin model. The nonlinear higher RPA is based on an equation nonlinear on eigenvectors ​​and includes many-particle-many-hole components in the creation operator of the excited states. We demonstrate the exact character of solutions analytically for the particle number $N$ = 2 and, numerically, for $N$ = 8. This finding indicates that the nonlinear higher RPA is equivalent to the exact Schr\"{o}dinger equation, which opens up new possibilities for realistic calculations in many-body problems.
\end{abstract}

\pacs{21.60.Jz, 71.10.-w}

\keywords{Random-phase approximation; Lipkin model; many-body problem.}

\maketitle

%\tableofcontents

\section{\label{sec:introduction}Introduction}
The random-phase approximation (RPA) \cite{Boh53a,Saw57b,Bro57,Wen57,Hub58b,Gel57} and its quasiparticle generalization (QRPA) \cite{Mar60,Arv60} have been, for a long time, very
important theoretical many-body methods in quantum chemistry, condensed matter physics and nuclear physics. 
Hence, it is natural to expect that an extension of the RPA will give a new more powerful method. Areas in need of more accurate methods of calculation include neutrino physics in connection with the search for the Majorana neutrino mass, constraints on which depend substantially on the nuclear matrix elements of neutrinoless double-$\beta$ decay \cite{Eng17}.

The RPA approach in its current formulation including  many refinements is an approximation 
which cuts off the excitations at the one-particle-one-hole (1p-1h) level. The extension of the RPA to include
also the 2p-2h excitations, so called the second RPA, has been 
investigated and used by many authors \cite{Suh62,Fan62a,Saw62,Dap65,Row68,Yan83,Tak88,Pap10,Gam16,Tak04,Sch16}. Currently the frontier of this approach
is the self-consistent second RPA \cite{Sch16}, which is still used only for schematic models (their equation is solved only approximately).
The aim of this work is the inclusion in the RPA framework of states with arbitrary order of particle-hole excitations from the ground state and the nonlinearity of the eigenequation \cite{Row68,Cat94,Cat96,Tak04,Sch16}, and our equations are solved exactly. It turns out that such extension reproduces the exact solutions of the Lipkin model \cite{Lip65}. 
In what follows, the extended RPA is referred to as the nonlinear higher RPA. 
We call the novel creation operator of the excited state the phonon operator for simplicity. Note, however, that {\it the boson commutation relation is not assumed.} 
%There are several ways to derive the RPA \cite{Rin80}, and a few of them give clues for the extension. Our 
%extension is based on the equation of motion method \cite{Row68} because one of us has experience of the 
%renormalized RPA \cite{Sim97}, which is also an extension of the RPA included in that method. 

In Sec.~\ref{sec:lipkin_model} we show the equations for applying our method to the Lipkin model, and the analytical (particle number of 2) and numerical (larger particle numbers) solutions are presented; these are the exact solutions. The result of the truncation approximation is also shown and compared with the shell model. In Sec.~\ref{sec:symmetry_breaking_basis} the formulation using the symmetry-breaking basis for the large interaction strength is discussed and numerically investigated. Section \ref{sec:summary} is devoted to summary. 

\section{\label{sec:lipkin_model}Application to Lipkin model}
\subsection{\label{subsec:formulation}Formulation}
The single-particle space of the Lipkin model consists of two fermion levels, each of which has an $N$-fold degeneracy.
The upper (lower) level has the energy of $\varepsilon/2$ ($-\varepsilon/2$). We  assume, without loss of generality,
that $N$ is even and equal to the particle number of the system. A parameter of $n = N/2$
is often used in this paper.
%Each single-particle state in the upper (lower) level is specified by indexes $1m$ ($0m$), $m=1, \cdots, N$. 
The system state in which all particles are in the lower level is denoted by $|\psi_0\rangle$.
The Hamiltonian of the Lipkin model is given by 
\begin{eqnarray}
H = \varepsilon J_z + \frac{V}{2}\left( J_+^2 + J_-^2 \right),
\end{eqnarray}
% P.1 in "Self-consistent 3rd RPA, Lipkin model, Aug. 24, 2016"
%For the definition of the operators $J_z$, $J_+$, and $J_-$, see Ref.~\cite{Lip65}. 
%\vspace{-16pt}
\begin{eqnarray}
J_z &=& \frac{1}{2}\sum_{m=1}^N \left( a^\dagger_{1m} a_{1m} - a^\dagger_{0m} a_{0m} \right), 
\label{eq:Jz}\\ \nonumber \\[0pt]
J_+ &=& \sum_{m=1}^N a^\dagger_{1m} a_{0m}, \ \ J_- = J_+^\dagger. \label{eq:J+-} 
\end{eqnarray}
The creation and annihilation operators of the fermion are denoted by $a^\dagger_{im}$ and $a_{im}$, 
($i=0$: lower level, $i=1$: upper level). Index $m$ distinguishes the degenerated states. $J_+$, $J_-$, and $J_z$ satisfy the following commutation relations:
\begin{eqnarray}
[ J_z, J_+ ] = J_+, \ \ [J_z, J_-] = -J_-, \ \ [J_+, J_-] = 2J_z. \label{eq:genr_comn}
\end{eqnarray}
$V$ is the strength of the interaction. Our purpose is to test our new method,
therefore we are interested in the space affected by the interaction \cite{Lip65}. For this reason,
the space relevant to us is spanned by state vectors $J_+^i|\psi_0\rangle$, $(i=0,\cdots,2n)$. This space
splits into two subspaces. One is the odd-order subspace with respect to $J_+$, and
another is the even-order subspace ($|\psi_0\rangle$ included).
%states as $a^\dagger_{1m} a_{0m^\prime}|\psi_0\rangle$, ($m\neq m^\prime$), are not treated in our 
%study. This is the basic idea of Lipkin et al.~in their paper which introduced this model. 
%The relevant solution space consists of two decoupled subspaces. One is spanned by 
%\begin{equation}
%\{J_+|\psi_0\rangle, J_+^3|\psi_0\rangle \cdots , J_+^{2n-1}|\psi_0\rangle\}, 
%\end{equation}
%and we call this the odd-order subspace. Another one is spanned by
%\begin{equation}
%\{|\psi_0\rangle, J_+^2|\psi_0\rangle \cdots , J_+^{2n}|\psi_0\rangle\},  
%\end{equation}
%and we call this the even-order subspace.

The following formulae can be derived from the commutation relations (\ref{eq:genr_comn}):
\begin{eqnarray}
&&J_z J_+^i | \psi_0 \rangle = f(i) J_+^i | \psi_0 \rangle, \ \ f(i) = i - \frac{N}{2}, \label{eq:fpsi0}\\ \nonumber \\[-10pt]
&&J_- J_+^i | \psi_0 \rangle = g(i) J_+^{i-1} | \psi_0 \rangle, \ \ g(i) = -i^2 + (N+1)i, \label{eq:gpsi0}
\label{eq:useful_formula_2}
\end{eqnarray}
% P.19 in "Self-consistent 3rd RPA, Lipkin model, Aug. 24, 2016"
for $0 \leq i \leq 2n$. We extend the definition of $f(i)$ and $g(i)$ to 
\begin{eqnarray}
f(i) = g(i) = 0, \ \ ( i<0,\ i>2n ),
\end{eqnarray}
% Check!
and introduce a function
\begin{eqnarray}
G(i) &=\left\{
\begin{array}{ll}
\displaystyle{\prod_{j=1}^i g(j)}, & (1 \leq i \leq N), \\ \\[-7pt]
1, & (i=0), \\ \\[-7pt]
0, & \textrm{otherwise},
\end{array}
\right. 
= \left\{
\begin{array}{ll}
\displaystyle{ \frac{i! N!}{(N-i)!} }, & (0 \leq i \leq N), \\ \\[-7pt]  
0, & \textrm{otherwise}. 
\end{array} 
\right.
\label{eq:G}
\end{eqnarray}
% Eq. (7.2) in "Reconstruct formulation for avoiding overflow, Nov. 13, 2016"

The phonon-creation operator $Q_k^\dagger$ ($k$ denotes an excited state) and the excitation energy $E_{k0}$ are determined by the equation of motion
\begin{eqnarray}
&&[ H, Q_k^\dagger ]|\Psi_0\rangle = E_{k0} Q_k^\dagger |\Psi_0\rangle . \label{eq:eq_motion}
\end{eqnarray}
The ground state $|\Psi_0\rangle$ is determined by
\begin{eqnarray}
Q_k |\Psi_0\rangle = 0, \label{eq:vac_cndn}
\end{eqnarray}
which we call the vacuum condition. $Q_k^\dagger|\Psi_0\rangle$ is the excited state, and  
its orthogonality to the ground state is guaranteed by the vacuum condition. 
%(We prefer this to ``killing condition" \cite{Jem13}.) 
The general framework is defined by Eqs.~(\ref{eq:eq_motion}) and (\ref{eq:vac_cndn}).

$Q_k^\dagger$ for the Lipkin model is set to
\begin{eqnarray}
Q_k^{o\dagger} &=& \sum_{l=1}^n \big( X^k_{2l-1} {\cal J}_+^{2l-1} + Y^k_{2l-1}{\cal J}_-^{2l-1} \big), \ \ \textrm{(odd-order subspace)}, \label{eq:Qko+} \\
Q_k^{e\dagger} &=& c_k + \sum_{l=1}^n \big( X^k_{2l} {\cal J}_+^{2l} + Y^k_{2l} {\cal J}_-^{2l} \big), \ \ \textrm{(even-order subspace)}, \label{eq:Qke+}
\end{eqnarray}
% Eqs. (2.1) and (12.1) in ''Reconstruct formulation for avoiding overflow, Nov. 13, 2016''
with 
\begin{align}
{\cal J}_\pm^i = J_\pm^i/\sqrt{G(i)}. 
\end{align}
Here, $c_k$ is a c-number determined together with $X_i^k$'s and $Y_i^k$'s by solving the equations. 
The denominator in ${\cal J}_\pm^i$ is introduced to avoid the overflow in the
numerical calculation of the Hamiltonian matrix elements for large $n$. 
The ground state can be written  
\begin{eqnarray}
|\Psi_0\rangle = \sum_{i=0}^n \beta_{2i}{\cal J}_+^{2i}|\psi_0\rangle.
\end{eqnarray}
% Eq. (7.3) in ''Reconstruct formulation for avoiding overflow, Nov. 13, 2016"
Equation (\ref{eq:eq_motion}) for the odd-order subspace can be cast into the matrix-vector form
\begin{eqnarray}
&&
\left(
\begin{array}{cc}
A^o & B^o \\
B^o & A^o 
\end{array}
\right)
\left(
\begin{array}{c}
\bm{X}^o_k \\
\bm{Y}^o_k
\end{array}
\right)
=E^o_{k0}
\left(
\begin{array}{cc}
U^o & O \\
O     & -U^o 
\end{array}
\right)
\left(
\begin{array}{c}
\bm{X}^o_k \\
\bm{Y}^o_k
\end{array}
\right)
, \label{eq:eigeneq}
\end{eqnarray}
\begin{eqnarray}
\bm{X}^o_k &=& ( X^k_1, X^k_3, \cdots, X^k_{2n-1} )^T, \\ \nonumber \\[-5pt]
\bm{Y}^o_k &=& ( Y^k_1, Y^k_3, \cdots, Y^k_{2n-1} )^T. 
\end{eqnarray}
$A^o$, $B^o$, and $U^o$ are $n \times n$ matrices, and the suffix $T$ stands for transpose. 
%The upper suffix $o$ of the sub-matrices and $E_{k0}^o$ indicates that those are the quantities of the odd-order subspace. 
The equation for the even-order subspace can be written analogously. With the abbreviation of the symmetric  double commutator
\begin{eqnarray}
[A,B,C] = \frac{1}{2}[[A,B],C] + \frac{1}{2}[A,[B,C]], 
\end{eqnarray}
the matrix elements in Eq.~(\ref{eq:eigeneq}) are defined by
\begin{eqnarray}
A_{ij}^{o} &=& \langle \Psi_0| [ {\cal J}_-^{2i-1}, H, {\cal J}_+^{2j-1} ] |\Psi_0\rangle, \label{eq:def_H++} \\ \nonumber \\[-5pt]
B_{ij}^{o} &=& \langle \Psi_0| [ {\cal J}_-^{2i-1}, H, {\cal J}_-^{2j-1} ] |\Psi_0\rangle, \\ \nonumber \\[-5pt]
U_{ij}^{o} &=& \langle \Psi_0 | [ {\cal J}_-^{2i-1} , {\cal J}_+^{2j-1} ] |\Psi_0\rangle ,
\end{eqnarray}
%{\cal U}_{ij}^{o--} &=& \langle \Psi_0 | [ \frac{(J_-^{2i-1})^\dagger}{\sqrt{G(2i-1)}} , \frac{J_-^{2j-1}}{\sqrt{G(2j-1)}} ] |\Psi_0\rangle, \\
%{\cal H}_{ij}^{o--} &=& \langle \Psi_0| [ \frac{(J_-^{2i-1})^\dagger}{\sqrt{G(2i-1)}}, H, \frac{J_-^{2j-1}}{\sqrt{G(2j-1)}} ] |\Psi_0\rangle, \\
%{\cal H}_{ij}^{o-+} &=& \langle \Psi_0| [ \frac{(J_-^{2i-1})^\dagger}{\sqrt{G(2i-1)}}, H, \frac{J_+^{2j-1}}{\sqrt{G(2j-1)}} ] |\Psi_0 \rangle, 
% Eqs. (4.2)-(4.5) in ''Reconstruct formulation for avoiding overflow, Nov. 13, 2016''
% Eqs.(17.2)-(17.9) in ''General formulation, Sep. 27, 2016''
% Eqs. (5.1)-(5.4) in ''Reconstruct formulation for avoiding overflow, Nov. 13, 2016''
% Eqs.(17.6)-(17.9) in ''General formulation, Sep. 27, 2016''
for $i,j=1,\cdots, n$.
The use of $|\Psi_0\rangle$ is the ultimate extension of the renormalized RPA \cite{Row68}.
The symmetric double commutator including $H$ is used for guaranteeing the symmetry of the Hamiltonian matrix: 
%The matrix elements of other block matrices can be obtained by the relations
%\begin{eqnarray}
%{\cal H}_{ij}^{o--} = {\cal H}_{ji}^{o++}, \ 
%{\cal H}_{ij}^{o-+} = {\cal H}_{ji}^{o+-} , \ 
%{\cal U}_{ij}^{o--} = -{\cal U}_{ji}^{o++}. 
%\end{eqnarray}
% Eqs. (5.6), (5.8), and (6.1) in "Reconstruct formulation for avoiding overflow, Nov. 13, 2016"
%We also have the symmetry relations of each block matrix
\begin{eqnarray}
%$
A_{ij}^{o} = A_{ji}^{o}, \ 
B_{ij}^{o} = B_{ji}^{o}.
%$
\end{eqnarray}
We also have 
\begin{eqnarray} 
U_{ij}^{o} = U_{ji}^{o}.
\end{eqnarray}
% Eqs. (5.8) and (6.2) in "Reconstruct formulation for avoiding overflow, Nov. 13, 2016"
The equations of the matrix elements and symmetry relations for the even-order subspace can be obtained analogously. Matrix elements of the even-order subspace are labeled with suffix $e$ (see below).  
The explicit equations of $A_{ij}^{o}$ and $A_{ij}^{e}$ can be obtained from
%\begin{widetext}
\begin{align}
\lefteqn{ \langle \Psi_0 | [ {\cal J}_-^k , H, {\cal J}_+^l ] |\Psi_0\rangle } \nonumber \\
&=\frac{\varepsilon}{2} \sum_{
{\scriptsize
\begin{array}{c}
i\!=\!0\\
(0\! \leq \! 2i \! + \! k \! - \! l \! \leq \! 2n)
\end{array}
}
}^n \frac{ \beta_{2i}\beta_{2i+k-l} }{ \sqrt{ G(k)G(l)G(2i)G(2i+k-l) } } \Big[ \{ 2f(2i+k) -f(2i) \nonumber \\
&-f(2i+k-l) \} G(2i+k) +\big\{ -f(2i-l+k) +2f(2i-l) -f(2i) \big\} \nonumber \\
&\times g(2i)g(2i-1)\cdots g(2i-l+1) G(2i-l+k) \Big] \nonumber \\ 
& +\frac{V}{4}\sum_{
{\scriptsize
\begin{array}{c}
i\!=\!0\\
(0\! \leq \! 2i\! +\! k \! - \! 2\! - \! l \! \leq \! 2n)
\end{array}
}
}^n \frac{ \beta_{2i}\beta_{2i+k-2-l} }{ \sqrt{ G(k)G(l)G(2i)G(2i+k-2-l) } } \big\{ 
G(2i+k) \nonumber \\
& -g(2i) g(2i-1) G(2i+k-2) - g(2i)g(2i-1)\cdots g(2i-l+1) G(2i-l+k) \nonumber \\
& +g(2i)g(2i-1)\cdots g(2i-l-1) G(2i-l-2+k) \big\} \nonumber \\ 
& +\frac{ V }{ 4 } \sum_{
{\scriptsize
\begin{array}{c}
i \! = \! 0\\
(0 \! \leq \! 2i \! + \! k \! + \! 2 \! - \! l \leq \! 2n)
\end{array}
}
}^n \frac{ \beta_{2i}\beta_{2i+k+2-l} }{ \sqrt{ G(k)G(l)G(2i)G(2i+k+2-l) } }  \big\{ G(2i+k+2) \nonumber \\
&+ g(2i)g(2i-1)\cdots g(2i-l+1) G(2i-l+2+k) \nonumber \\
&- g(2i+k)g(2i+k-1)\cdots g(2i+k-l+1) G(2i+k-l+2) \nonumber \\
&- g(2i+2)g(2i+1)\cdots g(2i+3-l) G(2i+2-l+k) \big\} , \label{eq:mat_el_A}
\end{align}
% Checked by comparing with scnrpaodd2.1.f90, Jun.7, 2017.
for $k,l=1,\cdots, 2n$. 
This equation is derived by making use of Eqs.~(\ref{eq:fpsi0})$-$(\ref{eq:G}). 
For $B_{ij}^{o}$ and $B_{ij}^{e}$, we can use
\begin{eqnarray}
\lefteqn{ \langle \Psi_0| [ {\cal J}_-^k, H, {\cal J}_-^l ] |\Psi_0\rangle } \nonumber \\
&&= \frac{V}{2} \! \sum_{
{\scriptsize
\begin{array}{c}
i\!=\!0\\
(0\! \leq \! 2i \! + \! k \! - \! 2 \! + \! l \! \leq \! 2n)
\end{array}
}
}^n \!\frac{ \beta_{2i}\beta_{2i+k-2+l} }{ \sqrt{G(k) G(l) G(2i)} } \bigg[ \big\{ g(2i+k)g(2i+k-1) - g(2i)g(2i-1) \nonumber \\
&& \hspace{10pt}+ g(2i+l)g(2i+l-1) \big\} \sqrt{ G(2i+l-2+k) } - \frac{G(2i+l+k)}{ \sqrt{G(2i+k-2+l)} } \bigg], \label{eq:mat_el_H}
\end{eqnarray}
% Eq. (4.2) in ''For avoiding overflow, Nov. 10, 2016"
and $U_{ij}^o$ and $U_{ij}^e$ are calculated by 
\begin{align}
\langle \Psi_0 | [ {\cal J}_-^k , {\cal J}_+^l ] | \Psi_0 \rangle =&\; 
\sum_{
{\scriptsize
\begin{array}{c}
i \! = \! 0\\
(0 \! \leq \! 2i \! + \! k \! - \! l \leq \! 2n)
\end{array}
}
}^n \frac{ \beta_{2i}\beta_{2i+k-l} }{ \sqrt{ G(k)G(l)G(2i)G(2i+k-l) } } \big\{ G(2i+k) \nonumber \\
& - G(2i) g(2i-l+k)g(2i-l+k-1)\cdots g(2i-l+1) \big\}. \label{eq:mat_el_U}
\end{align}
%\end{widetext}
% Checked by comparing with scnrpaodd2.1.f90, Jun.7, 2017.

The vacuum condition (\ref{eq:vac_cndn}) for $k$ belonging to the odd-order subspace yields
\begin{eqnarray}
&&\left(
\begin{array}{ccc}
L^k_{11} & \cdots & L^k_{1n} \\
 \vdots &  &\vdots \\
L^k_{n1} & \cdots & L^k_{nn}
\end{array}
\right) 
\left(
\begin{array}{c}
\beta_2/\beta_0 \\
\vdots \\
\beta_{2n}/\beta_0
\end{array}
\right) 
= 
\left(
\begin{array}{c}
-Y^k_{1} \\
\vdots \\
-Y^k_{2n-1}
\end{array}
\right), 
\label{eq:vac_cndn_odd}
\end{eqnarray}
% Eq. (11.0.1.2) in "Reconstruct formulation for avoiding overflow, Nov. 13, 2016" with \sqrt{G(i)}L_ij -> L_ij
% Eq. (33.2) in ''General formulation, Sep. 27, 2016''
i) $i=1,\cdots , n;\ j=i,\cdots ,n$, (upper triangle including the diagonal line)
\begin{eqnarray}
&&L^k_{ij} = 
\displaystyle{ 
X^k_{2j-2i+1} \sqrt{ \frac{G(i)G(2j)}{ G(2j-2i+1) } }\frac{1}{G(2i-1)}},
\label{eq:Lij_upper}
\end{eqnarray}
ii) $i=2,\cdots , n;\ j=1,\cdots ,i-1$, (lower triangle)
\begin{eqnarray}
&&L^k_{ij} = 
\displaystyle{
Y^k_{2i-2j-1} \sqrt{ \frac{G(i)}{G(2i-2j-1) G(2j)} }}. 
\label{eq:Lij_lower} 
%\label{eq:Lij}
\end{eqnarray}
% Eqs. (11.3) and (11.4) in "Reconstruct formulation for avoiding overflow, Nov. 13, 2016"
If $X^k_{2i-1}$'s and $Y^k_{2i-1}$'s are given, Eq.~(\ref{eq:vac_cndn_odd}) seems at first glance to indicate that $\beta_{2i}/\beta_0$'s depend on $k$.  Actually, the solution is independent of $k$. An analysis related to this property is shown below using the numerical result.  

For $k$ of the even-order subspace, the vacuum condition (\ref{eq:vac_cndn}) yields the following equations (note that there are $n+1$ equations):
\begin{eqnarray}
c_k &=& -\sum_{j=1}^n X^k_{2j}\frac{\beta_{2j}}{\beta_0}, \label{eq:alphak0_1} \\
c_k &=& -\sum_{j=l+1}^n X^k_{2j-2l}\frac{\beta_{2j}}{\beta_{2l}} 
\sqrt{ \frac{G(2j)}{ G(2j-2l) G(2l) } } -\sum_{j=0}^{l-1}Y^k_{-2j+2l}\frac{\beta_{2j}}{\beta_{2l}}
\sqrt{ \frac{G(2l)}{ G(2l-2j) G(2j) } }, \nonumber \\ \nonumber \\[-5pt] 
&& (l=1,\cdots , n-1), \label{eq:alphak0_3b}\\
c_k &=& -\sum_{j=0}^{n-1} Y^k_{-2j+2n}\frac{\beta_{2j}}{\beta_{2n}}
\sqrt{ \frac{ G(2n) }{ G(2n-2j) G(2j) } } \label{eq:alphak0_3}.
\end{eqnarray}
Apparently, any of these equations determines $c_k$, if $\beta_{2i}$'s, $X^k_{2i}$'s, and $Y^k_{2i}$'s are given. It is confirmed numerically below that $c_k$ is independent of the choice of the equation. 
For the treatment of the vacuum condition by previous papers, see, e.g., Refs.~\cite{Bal69,Duk90}. 
%We also note that it is possible to rewrite Eqs.~(\ref{eq:alphak0_3b}) and (\ref{eq:alphak0_3}) to a form of matrix-vector equation similar to Eq.~(\ref{eq:vac_cndn_odd}). 

The number of excited states is the same as that in the shell model (diagonalization of $H$ matrix represented by an orthonormal basis) as the phonon operators are constructed from the operators creating the orthonormal basis ($J_+^i$) and their hermite conjugates.   

\begin{table*}[!t]
  \caption{\label{tab:equations_N2} Eigenstates, wave functions, total energies, excitation energies, and phonon-creation operators obtained for $N=2$ by the nonlinear higher RPA. The wave functions and energies are identical to those obtained by the exact shell model. $E_{10}^o$ is given by Eq.~(\ref{eq:Eo10_N=2}). Each of the odd- and even- order subspaces has one excited state.}
%\begin{ruledtabular}
\centering
\begin{tabular}{ccc}
\hline
Eigenstate & Wave function & Total energy \\
\hline\\[-11pt]
Ground & $|\Psi_0\rangle = \frac{V}{\sqrt{2E_{10}^o(E_{10}^o-\varepsilon)}}\big( 1 -\frac{E_{10}^o-\varepsilon}{2V}J_+^2\big)|\psi_0\rangle$ & $-E_{10}^o$ \\ \\[-7pt]
Odd-order excited & $Q_{1}^{o\dagger}|\Psi_0\rangle = \frac{1}{\sqrt{2}}J_+|\psi_0\rangle$ & 0 
\\ \\[-7pt]
Even-order excited & 
$Q_{1}^{e\dagger}|\Psi_0\rangle = \frac{V}{\sqrt{2E_{10}^o(E_{10}^o+\varepsilon)}}\big(1+\frac{E_{10}^o+\varepsilon}{2V} J_+^2\big)|\psi_0\rangle$ & 
$E_{10}^o$ \\
\hline \\
\end{tabular}
%\end{ruledtabular}
%\vspace{20pt}
%\begin{ruledtabular}
\begin{tabular}{ccc}
\hline
Eigenstate &  Excitation energy & Phonon-creation operator \\
\hline\\[-11pt]
Ground & 0 &  \\
Odd-order excited & $E_{10}^o$ & 
$
\begin{array}{l}
Q_{1}^{o\dagger} = \frac{\sqrt{E_{10}^o}}{2\varepsilon}\Big(\frac{V}{|V|}\sqrt{E_{10}^o+\varepsilon}J_+ \\
\hspace{30pt}+\sqrt{E_{10}^o-\varepsilon}J_-\Big)
%= X_1^1{\cal J}_+ + Y_1^1{\cal J}_- \\
\end{array}
$
\\ \\[-7pt]
Even-order excited &  $E_{10}^e=2E_{10}^o$ & 
$
\begin{array}{l}
Q_{1}^{e\dagger} = \frac{V}{|V|}\Big( \frac{V}{2\varepsilon} + \frac{E_{10}^o+\varepsilon}{4\varepsilon}J_+^2 + \frac{E_{10}^o-\varepsilon}{4\varepsilon}J_-^2 \Big)
%= c_1 + X_2^1 {\cal J}_+^2 + Y_2^1 {\cal J}_-^2 \\
\end{array}
$ \\
\hline
\end{tabular}
%\end{ruledtabular}
\end{table*}

\subsection{\label{subsec:analytical_result}Analytical result}
The equations for the odd-order subspace with $N=2$ are considered analytically. Since there is only one excited state, we write
\begin{eqnarray}
Q_o^\dagger &=& \frac{1}{\sqrt{2}} \big( \alpha_+ J_+ + \alpha_- J_- \big).
%Q_e^\dagger &=&  \alpha_0 + \frac{1}{2} \big( \alpha_{2+} J_+^2 + \alpha_{2-} J_-^2 \big).
\end{eqnarray}
The vacuum condition (\ref{eq:vac_cndn}) gives
\begin{eqnarray}
\frac{\beta_2}{\beta_0} = -\frac{\alpha_-}{\alpha_+}, 
\label{eq:vac_cndn_N=2}
\end{eqnarray}
and the eigenequation (\ref{eq:eigeneq}) reads
\begin{eqnarray}
\left(
\begin{array}{cc}
{\cal H}_{++}^o & {\cal H}_{+-}^o \\
{\cal H}_{-+}^o  & {\cal H}_{--}^o
\end{array}
\right)
\left(
\begin{array}{c}
\alpha_+ \\
\alpha_-
\end{array}
\right)
= E_{10}^o
\left(
\begin{array}{cc}
{\cal U}_{++}^o & 0 \\
0                        & {\cal U}_{--}^o
\end{array}
\right)
\left(
\begin{array}{c}
\alpha_+ \\
\alpha_-
\end{array}
\right) . 
\label{eq:eigeneq_N=2}
\end{eqnarray}
The matrix elements are given by
\begin{eqnarray}
{\cal H}_{++}^o &=& \Big\{ \varepsilon - \varepsilon \Big(\frac{\alpha_-}{\alpha_+}\Big)^2 +2V\Big(\frac{\alpha_-}{\alpha_+}\Big) \Big\}\beta_0^2, \label{eq:H++}\\
{\cal H}_{+-}^o &=& -V \Big\{ 1 + \Big(\frac{\alpha_-}{\alpha_+}\Big)^2 \Big\}\beta_0^2, \label{eq:H+-}\\
{\cal U}_{++}^o &=& \Big\{ 1 - \Big(\frac{\alpha_-}{\alpha_+}\Big)^2 \Big\} \beta_0^2 , \label{eq:U++}
\end{eqnarray}
where Eq.~(\ref{eq:vac_cndn_N=2}) is used. 
From the above equations the following equation for $x\equiv \alpha_-/\alpha_+$ is obtained:
\begin{eqnarray}
V^2 x^6 +2\varepsilon V x^5 -V^2 x^4 -V^2 x^2 -2\varepsilon V x +V^2 = 0.
\end{eqnarray}
This algebraic equation has six solutions:
\begin{eqnarray}
x = \pm 1, \ \pm i, \ \frac{1}{V}\big( -\varepsilon \pm \sqrt{\varepsilon^2+V^2} \big).
\end{eqnarray}
We choose the physical solution
\begin{eqnarray}
\frac{\alpha_-}{\alpha_+} = \frac{1}{V}\big( -\varepsilon + \sqrt{\varepsilon^2+V^2} \big), \label{eq:a+/a-}
\end{eqnarray}
satisfying
\begin{eqnarray}
\lim_{V\rightarrow 0} \frac{\alpha_-}{\alpha_+} = 0.
\end{eqnarray}
Equations (\ref{eq:vac_cndn_N=2}) and (\ref{eq:a+/a-}) give the ratio of the components of the exact ground state. The excitation energy is obtained
\begin{eqnarray}
E_{10}^o = \sqrt{\varepsilon^2 + V^2}, \label{eq:Eo10_N=2}
\end{eqnarray}
which is identical to the exact one. The wavefunction $Q_o^\dagger|\Psi_0\rangle$ is equal to $ \frac{1}{\sqrt{2}}J_+|\psi_0\rangle$. 
It is possible to reproduce the exact solutions of the even-order subspace without high-order algebraic equation by using $\beta_0$ and $\beta_2$ obtained by the odd-order subspace calculation. 
The analytical solutions for $N=2$ are summarized in Table \ref{tab:equations_N2}. It is possible to confirm the explicit equations of $Q_{1}^{o\dagger}|\Psi_0\rangle$ and $Q_{1}^{e\dagger}|\Psi_0\rangle$ as well as $Q_{1}^o|\Psi_0\rangle=0$ and $Q_{1}^e|\Psi_0\rangle=0$ using the equations of this table. 

If the equations of the self-consistent second RPA \cite{Tak04,Jem13,Sch16} with $c_k$ are solved exactly for $N=2$, our result should be obtained. 
The vacuum condition for the Lipkin model is solved in Ref.~\cite{Duk90}. The self-consistent RPA \cite{Jem13,Del05} reproduces the ground and first excited (the odd-order subspace) states for $N=2$. The studies of Refs.~\cite{Duk90,Del05} do not obtain the excited states in the even-order subspace because the phonon operators in these studies have only the 1p-1h components. That of Ref.~\cite{Jem13} includes the 2p-2h components in the phonon operator but does not obtain the exact even-order excited state for $N=2$ because $c_k$ is not used. 
The excitation energy of the RPA (with the 1p-1h phonon operator and the $A^o$, $B^o$, and $U^o$ calculated with $|\psi_0\rangle$) is $\sqrt{\varepsilon^2-V^2}$ (odd-order subspace), therefore, $V=\varepsilon$ is the breaking point of the RPA. 
It is seen analytically from the table that this problem does not occur in the nonlinear higher RPA. 

\subsection{\label{subsec:numerical_result}Numerical result}
\begin{figure}[!t]
\centering\includegraphics[width=7.0cm]{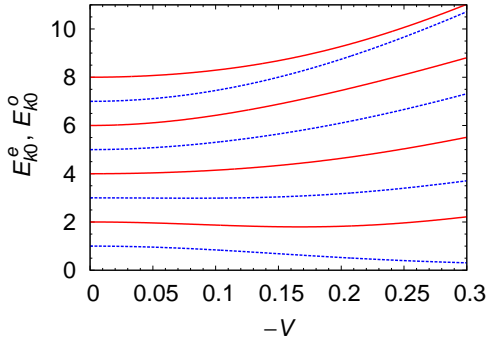}
\caption{ \label{fig:Ek0_odd} (Color online) $E_{k0}^e$ (solid line) and $E_{k0}^o$ (dashed line) as a function of $-V$ for $N=8$ and $\varepsilon=1$. The breaking point of the RPA is $V=-0.143$.}
\end{figure}
%\begin{figure}
%\includegraphics[width=7.5cm]{vw}
%\caption{ \label{fig:Ek0_even} (Color online) $E_{k0}^e$ as a function of $-V$.}
%\end{figure}
\begin{figure}[!t]
\centering\includegraphics[width=7.0cm]{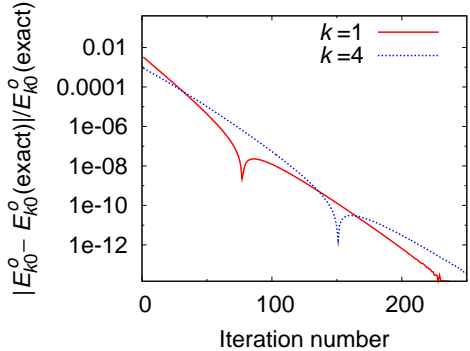}
\caption{ \label{fig:iw1_10} (Color online) Relative numerical error of $E_{k0}^o$ in iteration process. 
$E_{k0}^o(\textrm{exact})$ is the value obtained by the exact shell model ($V=-0.15)$: 
0.68405768907156 ($k=1$) and  
8.0011654731795 ($k=4$). 
The indicated $k$ is the excited-state number in the odd-order subspace. }
\end{figure}
For the initial calculation of the matrix elements in the nonlinear higher RPA equations we used an ansatz for the nonlinear higher RPA ground state as follows \cite{Sim00}:
\begin{align}
\frac{1}{\cal N} \exp \left[ \frac{Y}{X}J_+^2\right]|\psi_0\rangle,
\end{align}
where ${\cal N}$ is the normalization factor, with a small arbitrary value of $Y/X$ for generating the initial $\beta_{2i}$'s by expanding the exponential operator function [($J_+)^i|\psi_0\rangle=0$ for $i>2n$] . As $Y/X$ is small, $\beta_2$, $\beta_4, \cdots$ are very small. 
%We added a reference of \v{S}imkovic et al., PRC {\bf 61}, 044319 (2000) for this initial guess of the nonlinear higher RPA ground state. 
This initial guess was applied to the calculation for a small $V$, and for a slightly larger $V$ the solution for the slightly smaller $V$ was used as the initial guess. The solutions with increasing $V$ were obtained by repeating this manner. 

The initial $\beta_{2i}$'s are used for calculating the matrix elements  entering Eq.~(\ref{eq:eigeneq}) through Eqs.~(\ref{eq:mat_el_A})$-$(\ref{eq:mat_el_U}), and $X^k_{2i-1}$'s and $Y^k_{2i-1}$'s are obtained for all $k$ by solving Eq.~(\ref{eq:eigeneq}).  In this procedure we use a technique to diagonalize  $U^{o-1}(A^o+B^o)U^{o-1}(A^o-B^o)$ and obtain eigenvalue of $E_{k0}^2$ \cite{Rin80}. 
%(for the normalization see below). 
The orthonormalization condition is 
\begin{align}
\delta_{kk^\prime} &=\,
\langle \Psi_0 | Q_{k^\prime}^o Q_k^{o\dagger} | \Psi_0 \rangle = 
\langle \Psi_0 | [ Q_{k^\prime}^o, Q_k^{o\dagger} ] | \Psi_0 \rangle = ( \bm{X}^{oT}_{k^\prime}, \bm{Y}^{oT}_{k^\prime} ) 
\left(
\begin{array}{cc}
U^o & O \\
O & -U^o
\end{array}
\right) 
\left(
\begin{array}{c}
\bm{X}^o_k \\
\bm{Y}^o_k
\end{array}
\right) ,
\end{align}
and the one for the even-order subspace can be written in the same way. 
Then, these $X^k_{2i-1}$'s and $Y^k_{2i-1}$'s are input to Eq.~(\ref{eq:vac_cndn_odd}), and $\beta_{2i}/\beta_0$'s are obtained. Equation (\ref{eq:vac_cndn_odd}) with $k=1$ is used (here is the arbitrarity of the choice of $k$, as mentioned above, see also below). The component  $\beta_0$ is determined by the normalization of $|\Psi_0\rangle$;  
\begin{align}
\beta_0 = \frac{1}{\sqrt{1+\sum_{i=1}^n (\beta_{2i}/\beta_0)^2}}. \label{eq:beta0}
\end{align}
The $\beta_{2i}$'s obtained from Eqs.~(\ref{eq:vac_cndn_odd}) and (\ref{eq:beta0}) are input to Eq.~(\ref{eq:eigeneq}) through Eqs.~(\ref{eq:mat_el_A})$-$(\ref{eq:mat_el_U}), and this procedure is iterated until the convergence is obtained. 
The converged $\beta_{2i}$'s are input to the eigenequation of the even-order subspace corresponding to Eq.~(\ref{eq:eigeneq}), and $X^k_{2i}$'s and $Y^k_{2i}$'s are obtained for all $k$; the iteration is not necessary at this stage. 
The $\beta_{2i}$'s, $X^k_{2i}$'s, and $Y^k_{2i}$'s are input to Eq.~(\ref{eq:alphak0_1}) (again there is an arbitrarity of the choice of the equation as mentioned above), and $c_k$ is determined for all $k$. 

%Equations (\ref{eq:eigeneq}) and (\ref{eq:vac_cndn_odd}) with $k=1$ are solved by iteration. The obtained $\beta_{2i}$'s are input to the eigenequation of the even-order subspace; iteration is not necessary for solving this equation. Finally, $c_k$ is determined by Eq.~(\ref{eq:alphak0_1}). 
In the numerical calculation, $\varepsilon$ is set equal to 1, and $N=2n=8$ is used. 
%Figure \ref{fig:Ek0_odd} shows  $E_{k0}^o$, $(k=1,2,\cdots ,n)$. 
The spectrum is shown in Fig.~\ref{fig:Ek0_odd} as a function of $-V$. 
We assign the excited-state numbers as $E_{10}^e < E_{20}^e < E_{30}^e < E_{40}^e$ and analogously for the odd-order subspace. 
The RPA breaking occurs at $V = -1/(N-1) = -0.143$, and we calculated up to about twice this strength. 
The accuracy check with $V=-0.15$ is shown by 
Fig~\ref{fig:iw1_10},  which shows that our calculation reproduces the exact result of the shell model with no truncation of the wavefunction space. Other excitation energies have the accuracies in the same range. We also confirmed that the components of the ground and excited states are equal to those of the exact calculation. 
$V=-0.15$ is always used in the analysis of this section.

\begin{figure}[!t]
\centering\includegraphics[width=7.0cm]{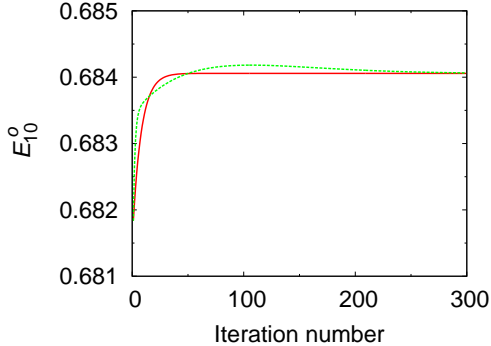}
\caption{ \label{fig:iw_vack1_2} (Color online) The convergence of $E_{10}^o$ obtained using 
$Q_{k=1}^o|\Psi_0\rangle=0$ (solid line) and $Q_{k=2}^o|\Psi_0\rangle=0$ (dashed line).}
\end{figure}

\begin{figure}[t]
\centering\includegraphics[width=7.0cm]{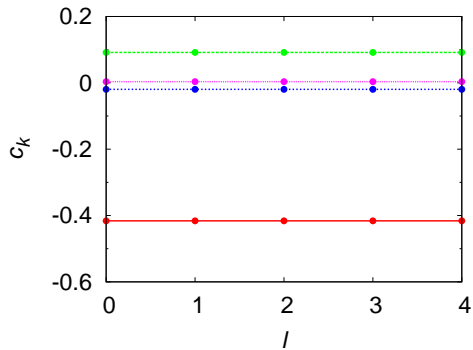}
\caption{ \label{fig:l_alpha} (Color online) Obtained $c_k$ using Eqs.~(\ref{eq:alphak0_1})$-$(\ref{eq:alphak0_3}) distinguished by $l$. Equation  $(\ref{eq:alphak0_1})$ [(\ref{eq:alphak0_3})] corresponds to $l=0$ ($n$). The order of $c_k$ is $|c_1|>|c_2|>|c_3| >|c_4|$.}
\end{figure}

The determination of $|\Psi_0\rangle$ is most sensitive to the lowest excited states through the vacuum condition. 
This is shown by Fig.~\ref{fig:iw_vack1_2}, which illustrates the convergence of the self-consistent calculation to the exact result obtained using the vacuum conditions with $k=1$ and 2. 
The lowest excited state $k=1$ was used for the vacuum condition of the calculation of Fig.~\ref{fig:iw1_10} because of this sensitivity. 
Figure \ref{fig:l_alpha} shows that $c_k$'s are obtained independently of the choice of the equation. This check is satisfactory for all $l$ specifying the equation. 

%\begin{figure}[!t]
%\includegraphics[width=8.5cm]{iw_vack1_2_vack_betag2}
%\caption{ \label{fig:iw_vack1_2} (Color online) The convergence of $E_{10}^o$ obtained using 
%$Q_1^o|\Psi_0\rangle=0$ (solid line) and $Q_2^o|\Psi_0\rangle=0$ (dashed line).  
%Inset: the output $\beta_{2i}$'s of one iteration with the input exact $\beta_{2i}$'s. The number $k$ indicates that $Q_k^o|\Psi_0\rangle=0$ is used. The 
%order of $\beta_{2i}$'s is $\beta_{0}>\beta_2>\cdots >\beta_{2n}$. Some $\beta_{2i}$'s are negligible and not seen.}
%\end{figure}
%\begin{figure}[t]
%\includegraphics[width=8.5cm]{vack_betag2}
%\caption{ \label{fig:vack_betag2} (Color online) The output $\beta_{2i}$'s of one iteration with the exact input $\beta_{2i}$'s. The number $k$ indicates that $Q_k^o$ is used for the vacuum condition. The order of $\beta_{2i}$'s is $\beta_{0}>\beta_2>\cdots >\beta_{2n}$. Some $\beta_{2i}$'s are negligible and not seen.}
%\end{figure}
\begin{figure}[t]
\centering
\begin{minipage}{1.0\textwidth}
\includegraphics[width=6.4cm]{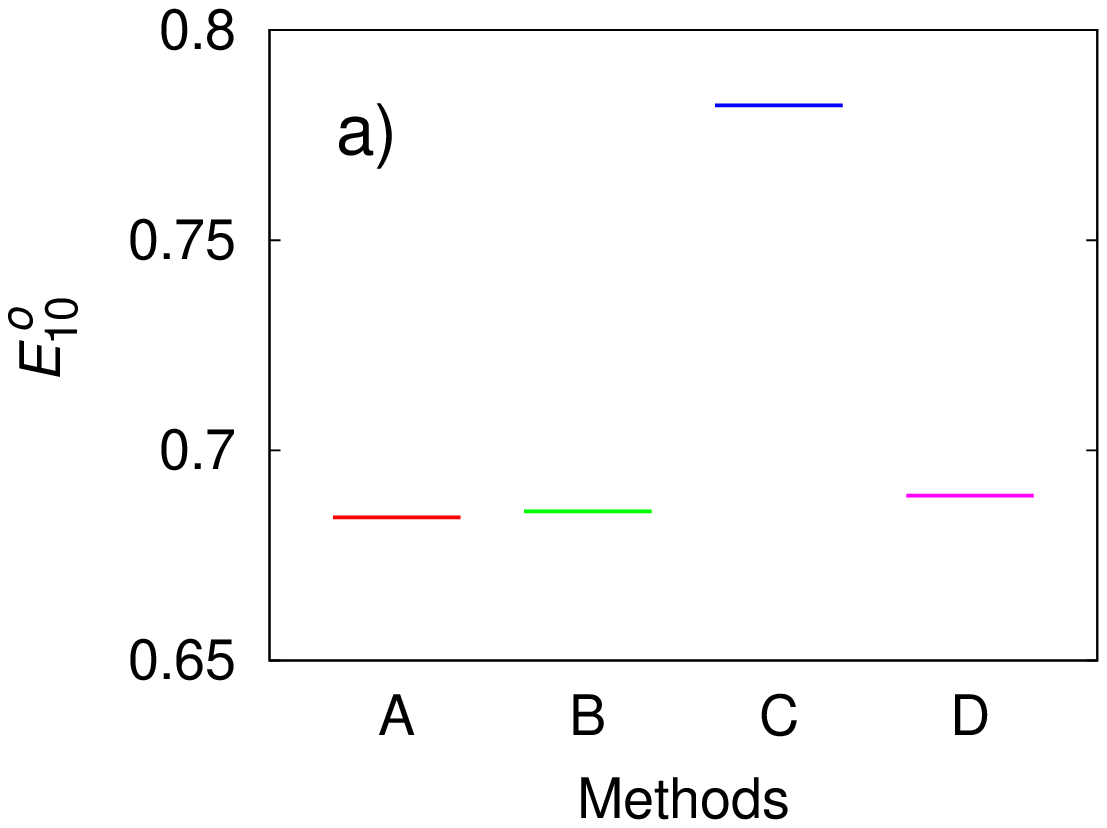}
\hspace{-9pt}
\includegraphics[width=6.4cm]{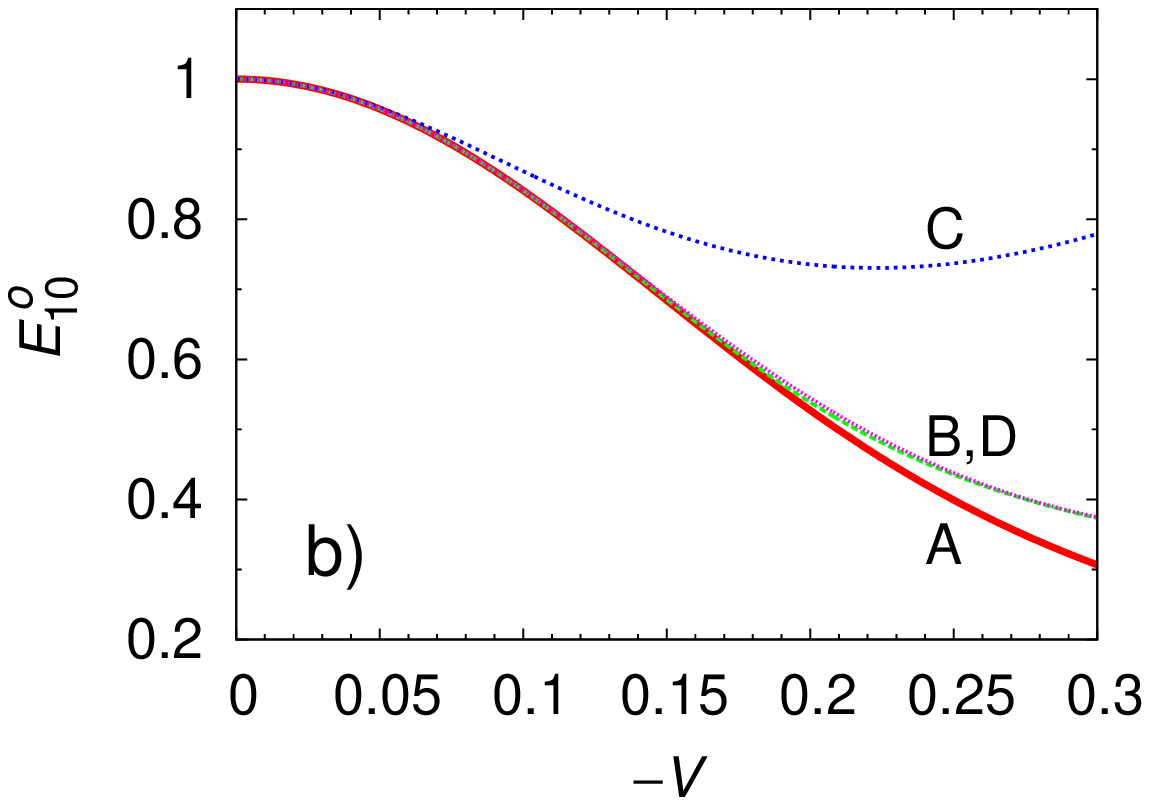}
\end{minipage}
\caption{ \label{fig:comp_methods_N8} (Color online) a) $E_{10}^o$ at $V=-0.15$ by four methods. A is the exact result, and the others are those obtained under the truncation of the matrices. B shows the nonlinear higher RPA, and C and D are the results of the shell model with different truncations. See text for detail. b) $V$ dependence of $E_{10}^o$. }
\end{figure}
\begin{figure}[t]
\begin{minipage}{1.0\textwidth}
\centering
\includegraphics[width=6.3cm]{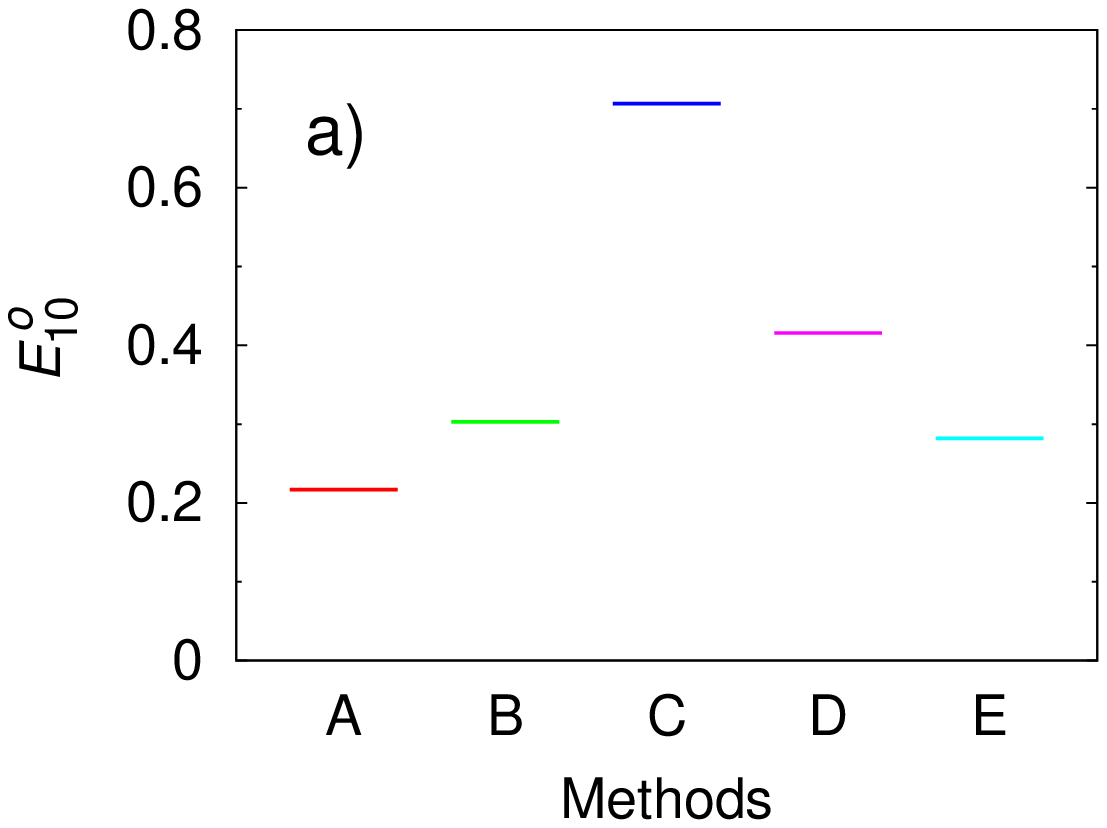}
\hspace{-9pt}
\includegraphics[width=6.3cm]{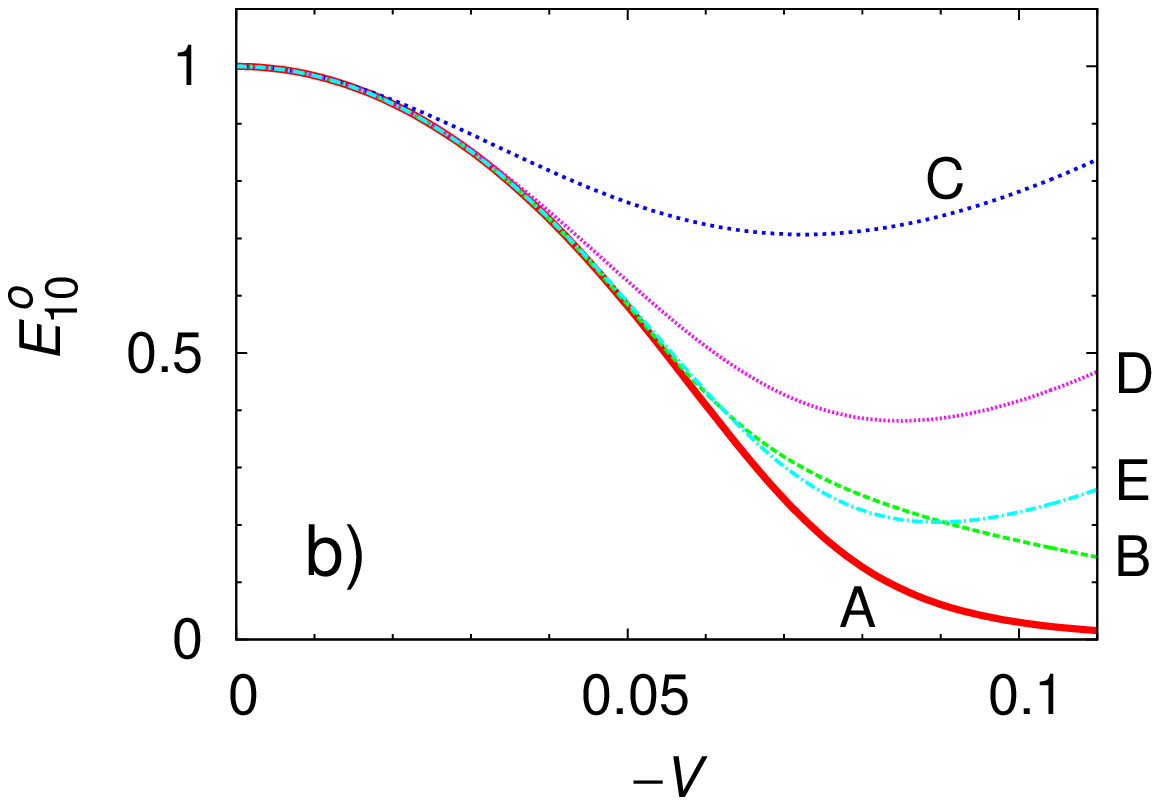}
\end{minipage}
\caption{ \label{fig:comp_methods} (Color online) The same comparison as Fig.~\ref{fig:comp_methods_N8} but for $N=20$. As that figure, A is the exact result, B shows that of the nonlinear higher RPA with a truncation, and the others are the results of the shell model with different truncations. See text for detail. a) $E_{10}^o$ at $V=-0.072$. b) $V$ dependence of $E_{10}^o$.}
\end{figure}

\subsection{\label{subsec:truncation_approximation}Truncation approximation}
For realistic calculations one cannot avoid approximation in any many-body approach. 
We compare the quality of the approximation under the truncation of the matrices between the nonlinear higher RPA and the shell model. 
As mentioned above, we treat $A^{o}$, $B^{o}$, and $U^{o}$ and those of the even-order subspace for solving the equations. Thus, the dimension of these matrices and that of the $H$ matrix of the shell model are referred to in the comparison. 
We performed calculations with $N=8$ and $V=-0.15$ (Fig.~5a).
This figure illustrates $E_{10}^o$ of four methods. Method A is the exact shell model  (the dimension of the matrix is $d=4$). Method B is the nonlinear higher RPA with $d=2$ ($n$ is still 4). Methods C and D are the shell model with different truncations: $d=2$ (C) and 3 (D). The results of methods B and C show that the nonlinear higher RPA is much better than the shell model with the same $d$. The shell model  needs $d=3$  for obtaining the quality of our method with $d=2$. This is understood from the feature of the nonlinear higher RPA. When the highest order of $J_+$ in $Q_k^{o\dagger}$ is $2i-1$, that in the wavefunction is $4i-1$.
The corresponding order of the shell model is $2i-1$. Thus, this advantage of the nonlinear higher RPA is expected well in the realistic calculations. Panel b of Fig.~\ref{fig:comp_methods_N8} shows the $V$ dependence of $E_{10}^o$. The truncation approximation B and D are very good up to $V \simeq -0.2$. 

We made the same comparison for $N=20$ ($n=10$), $V$ = $-$0.072, and the result is shown in panel a of Fig.~\ref{fig:comp_methods}. 
Method A is the exact calculation ($d$ = 10), and method B is the nonlinear higher RPA with $d$ = 2 ($n$ is still 10). Methods C, D, and E are the shell model with $d$ = 2 (C), 3 (D), and 4 (E). The same tendency as Fig.~\ref{fig:comp_methods_N8} is seen. But it is clear in this example with larger $N$ that the nonlinear higher RPA calculation with half the matrix size of the shell model calculation has the comparable approximation quality with the shell model calculation as is expected from the above argument. 
The method B ($d=2$) is equivalent to the self-consistent (extended) second RPA \cite{Sch16} (in our calculation the truncation is the only approximation). 
The $V$ dependence of $E_{10}^o$ is shown by panel b of Fig.~\ref{fig:comp_methods}. The truncation approximation B and E are good up to $V\simeq -0.06$. 

\section{\label{sec:symmetry_breaking_basis}Representation with symmetry-breaking basis}
In the previous section, the vacuum condition for the odd-order subspace determines the $\beta$'s, and that for the even-order subspace is used for determining $c_k$. In this section, we show that the ground state can also be determined without symmetry. In fact, the formulation with no symmetry has an opportunity to use for the Lipkin model because, if $|V|$ is large, the Hartree-Fock (HF) ground state breaks the symmetry. 

\subsection{\label{subsec:HF} Hartree-Fock basis}
We show the result of the HF approximation for the Lipkin model without derivation.
The HF equation is given by 
\begin{align}
\bigg(
\begin{array}{cc}
h_{11}^{HF} & h_{12}^{HF} \\
h_{21}^{HF} & h_{22}^{HF}
\end{array}
\bigg)
\bigg(
\begin{array}{c}
x_1^{i} \\
x_2^{i}
\end{array}
\bigg) = \epsilon_i^{HF}
\bigg(
\begin{array}{c}
x_1^{i} \\
x_2^{i}
\end{array}
\bigg), \ \ (i=1,2) \label{eq:HF_eq},   
\end{align}
\begin{align}
h_{11}^{HF} &= \frac{\varepsilon}{2}, \ \ 
h_{12}^{HF} = V(N-1)\langle \tilde{\psi}_0 | a_{1m}^\dagger a_{0m} | \tilde{\psi}_0 \rangle, \label{eq:h_12_HF} \\
h_{21}^{HF} &= h_{12}^{HF}, \ \ 
h_{22}^{HF} = -h_{11}^{HF}, 
\end{align}
where $|\tilde{\psi}_0\rangle$ denotes the HF ground state, and it is identical to $|\psi_0\rangle$ if $|V|$ is small (see below). 
The eigenvectors and eigenvalues are as follows:  
\begin{align}
x_1^1 &= \frac{h_{12}^{HF}}{ \sqrt{ {h_{12}^{HF} }^2 + ( \epsilon_1^{HF} - h_{11}^{HF} )^2 } }, \ \ 
x_2^1  = \frac{\epsilon_1^{HF} - h_{11}^{HF}}{ \sqrt{ {h_{12}^{HF} }^2 + ( \epsilon_1^{HF} - h_{11}^{HF} )^2 } }, \nonumber \\ 
x_1^2 &= \frac{h_{12}^{HF}}{ \sqrt{ {h_{12}^{HF} }^2 + ( \epsilon_2^{HF} - h_{11}^{HF} )^2 } }, \ \ 
x_2^2  = \frac{\epsilon_2^{HF} - h_{11}^{HF}}{ \sqrt{ {h_{12}^{HF} }^2 + ( \epsilon_2^{HF} - h_{11}^{HF} )^2 } }, 
\end{align}
\begin{align}
\epsilon_1^{HF} &= \sqrt{ \frac{\varepsilon^2}{4} + \{ V(N-1)\langle \tilde{\psi}_0 | a_{1m}^\dagger a_{0m} | \tilde{\psi}_0 \rangle \}^2 }, \label{eq:HF_energy_1} \ \ 
\epsilon_2^{HF}  = -\epsilon_1^{HF}. 
\end{align}
The annihilation operators $\alpha_{1m}$ and $\alpha_{0m}$ of the particles of the HF eigenstate are derived by 
\begin{align}
\bigg(
\begin{array}{c}
\alpha_{1m} \\
\alpha_{0m}
\end{array}
\bigg) = 
\bigg(
\begin{array}{cc}
x_1^1 & x_2^1 \\
x_1^2 & x_2^2 
\end{array}
\bigg)
\bigg(
\begin{array}{c}
a_{1m} \\
a_{0m}
\end{array}
\bigg), 
\end{align}
and these operators satisfy
\begin{align}
\alpha_{1m}|\tilde{\psi}_0 \rangle = \alpha_{0m}^\dagger|\tilde{\psi}_0 \rangle = 0. \label{eq:vacuum_condition_alpha}
\end{align}

\begin{figure}[t]
\centering\includegraphics[width=7.0cm]{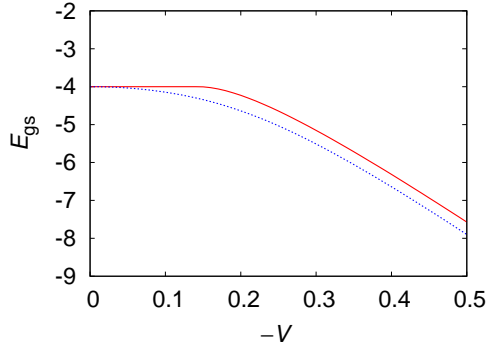}
\caption{ \label{fig:HF_gs_energy} (Color online) Ground-state energy of HF approximation (solid) and exact calculation (dashed) as a function of $-V$. Used are $\varepsilon=1$ and $N=8$. }
\end{figure}

The equation for $\langle a_{1m}^\dagger a_{0m} \rangle$ $\equiv$ $\langle \tilde{\psi}_0| a_{1m}^\dagger a_{0m} |\tilde{\psi}_0\rangle$ can be derived:
\begin{align}
&\langle a_{1m}^\dagger a_{0m} \rangle \bigg\{ \sqrt{ \frac{ \varepsilon^2 }{ 4 } +V^2(N-1)^2\langle a_{1m}^\dagger a_{0m}\rangle^2 }
+ \frac{\varepsilon}{2} \bigg\} \nonumber \\
&\times \bigg\{ 2\sqrt{ \frac{ \varepsilon^2 }{ 4 } +V^2(N-1)^2\langle a_{1m}^\dagger a_{0m}\rangle^2 } + V(N-1) \bigg\} = 0, \label{eq:secular_eq}
\end{align} 
and we eventually obtain
\begin{align}
&\langle a_{1m}^\dagger a_{0m} \rangle = 0, \ \ \Big( |V| \leq \frac{ \varepsilon }{N-1} \Big) , \label{eq:HF_solution_exp_val_1} \\ 
&\langle a_{1m}^\dagger a_{0m} \rangle = \frac{1}{2V(N-1)} \sqrt{ V^2(N-1)^2 - \varepsilon^2 }, \ \ \Big( |V| \geq \frac{ \varepsilon }{N-1} \Big). \label{eq:HF_solution_exp_val_2}
\end{align}
The HF solutions are determined by inserting Eq.~(\ref{eq:HF_solution_exp_val_1}) or (\ref{eq:HF_solution_exp_val_2}) to Eqs.~(\ref{eq:h_12_HF}) and (\ref{eq:HF_energy_1}). 
The HF ground state energy $E_{gs}^{HF}$ $=$ $\langle \tilde{\psi}_0 | H | \tilde{\psi}_0 \rangle$ is found to be 
\begin{align}
&E_{gs}^{HF} = -\frac{N\varepsilon}{2}, \ \ \Big(|V| \leq \frac{\varepsilon}{N-1}\Big), \label{eq:HF_gs_energy_1}\\
&E_{gs}^{HF} = \frac{1}{4} \frac{N}{N-1} \frac{\varepsilon^2}{V} + \frac{1}{4} N(N-1)V, \ \  \Big(|V| \geq \frac{\varepsilon}{N-1}\Big), \label{eq:HF_gs_energy_2}
\end{align}
and its behavior is drawn in Fig.~\ref{fig:HF_gs_energy}. For $|V|$ $>$ $\varepsilon/(N-1)$, the HF solution breaks the parity symmetry of the order with respect to $J_+$ ($\langle J_+ \rangle \neq 0$). As seen from Eq.~(\ref{eq:secular_eq}), the two HF solutions belong to the different branches. That is, one solution cannot be obtained from another one by changing the parameters. 

\subsection{\label{subsec:nhRPA_symmetry-breaking} Nonlinear higher RPA with symmetry-breaking basis}
In this section the notation of $|\tilde{\psi}_0\rangle$ is used for the symmetry-breaking HF ground state [$|V|$ $>$ $\varepsilon/(N-1)$]. 
We introduce the operators using the symmetry-breaking basis;
\begin{align}
\tilde{J}_z &= \frac{1}{2}  \sum_{m=1}^N
\left( \alpha^\dagger_{1m} \alpha_{1m} - \alpha^\dagger_{0m} \alpha_{0m} \right), \label{eq:tilde_Jz}\\
\tilde{J}_+ &= \sum_{m=1}^N \alpha^\dagger_{1m} \alpha_{0m},  \ \ \tilde{J}_- = \tilde{J}_+^\dagger . \label{eq:tilde_J+_J-}
\end{align} 
$\tilde{J}_z$, $\tilde{J}_+$, and $\tilde{J}_-$ satisfy the same commutation relations as those for $J_z$, $J_+$, and $J_-$.  
In this representation, the Hamiltonian is expressed 
\begin{align}
H =&\; \tilde{h}_z \tilde{J}_z + \tilde{h}_{+-} ( \tilde{J}_+ + \tilde{J}_- ) + \tilde{h}_{z2} \tilde{J}_z^2 
+ \tilde{h}_{+2-2} ( \tilde{J}_+^2 + \tilde{J}_-^2 ) + \tilde{h}_{+zz-} ( \tilde{J}_+ \tilde{J}_z + \tilde{J}_z \tilde{J}_- ) \nonumber \\
&+ \tilde{h}_{-+} \tilde{J}_- \tilde{J}_+ , \label{eq:H_symmetry_breaking}
\end{align}
\begin{align}
&\tilde{h}_z = \frac{1}{ |V|(N-1)^2 } \Big\{ \varepsilon^2 \Big(N-1+\frac{1}{2}\frac{V}{|V|}\Big) 
- \frac{1}{2} V^2 \frac{V}{|V|} (N-1)^2 \Big\}, \\
&\tilde{h}_{+-} = \frac{ \sqrt{ V^2 (N-1)^2 - \varepsilon^2 } }{ |V| (N-1)^2 } \frac{\varepsilon}{2} \Big( N - 1 - \frac{V}{|V|} \Big) , \\
&\tilde{h}_{z2} = \frac{ V^2 (N-1)^2 - \varepsilon^2 }{ V (N-1)^2 } , \\
&\tilde{h}_{+2-2} = \frac{ V^2 (N-1)^2 + \varepsilon^2 }{ 4V (N-1)^2 }, \\
&\tilde{h}_{+zz-} = -\varepsilon \frac{ \sqrt{ V^2 (N-1)^2 - \varepsilon^2 } }{ V (N-1)^2 }, \\
&\tilde{h}_{-+} = -\frac{ V^2 (N-1)^2 - \varepsilon^2 }{ 2V (N-1)^2 }.
\end{align} 
As seen from the linear terms of $\tilde{J}_+$ and $\tilde{J}_-$ in Eq.~(\ref{eq:H_symmetry_breaking}), the odd-order subspace $\{ \tilde{J}_+^{2i+1}|\tilde{\psi}_0\rangle \}$ and even-order one $\{ \tilde{J}_+^{2i}|\tilde{\psi}_0\rangle \}$ are not decoupled. 
The eigenequation of the nonlinear higher RPA in this representation can be derived analogously to Eq.~(\ref{eq:eigeneq}). Its extension to the symmetry-breaking formulation is straightforward, and we omit the explicit equations. 

The expression of the vacuum condition (\ref{eq:vac_cndn}) without using the symmetry is obtained 
\begin{align}
&\left(
\begin{array}{cccccc}
\tilde{c}_k                                  & \tilde{L}^k_{12}         &                              & \cdots                    &           & \tilde{L}^k_{1N} \\
\tilde{L}^k_{21} & \tilde{c}_k                                 & \tilde{L}^k_{23} & \cdots                    &           & \tilde{L}^k_{2N} \\
\tilde{L}^k_{31} & \tilde{L}^k_{32} & \tilde{c}_k                        &                              &           & \vdots \\
\vdots                             &                                        &                             &                              &  \ddots & \tilde{L}^k_{N-1\ N}\\
\tilde{L}^k_{N1} & \cdots                              &                             &                              & \tilde{L}^k_{N\ N-1} & \tilde{c}_k 
\end{array}
\right)
\left(
\begin{array}{c}
\tilde{\beta}_1/\tilde{\beta}_0 \\
\tilde{\beta}_2/\tilde{\beta}_0 \\
\vdots \\
\\
\tilde{\beta}_{N}/\tilde{\beta}_0
\end{array}
\right) 
= 
\left(
\begin{array}{c}
-\tilde{Y}^k_1 \\
-\tilde{Y}^k_2 \\
\vdots \\
\\
-\tilde{Y}^k_{N}
\end{array}
\right) , \label{eq:vacuum_condition_comp}
\end{align}
\begin{align}
\tilde{L}^k_{ij} &= \tilde{X}^k_{j-i} \sqrt{ \frac{G(j)}{G(j-i)G(i)} }, \ \ (i<j), \\
\tilde{L}^k_{ij} &= \tilde{Y}^k_{i-j} \sqrt{ \frac{G(i)}{G(i-j)G(j)} }, \ \ (i>j),  
\end{align}
\begin{align}
\tilde{c}_k = -\sum_{i=1}^N \tilde{X}^k_i \frac{ \tilde{\beta}_i}{ \tilde{\beta}_0 }, \label{eq:eq_c_k}
\end{align}
where $\tilde{\beta}$'s are the components of the ground state
\begin{align}
|\Psi_0\rangle = \sum_{l=0}^N \tilde{\beta}_l \frac{ \tilde{J}_+^l }{\sqrt{ G(l) }} | \tilde{\psi}_0 \rangle , 
\end{align}
and $\tilde{X}_i^k$'s and $\tilde{Y}_i^k$'s are the amplitudes obtained from the eigenequation in the symmetry-breaking representation (omitted). 
We applied this symmetry-breaking formulation for $N=8$, $V=-0.5$ and confirmed that the exact solutions are obtained. 
For calculating $c_k$, $\tilde{\beta}_i$'s of the previous cycle in the iteration can be used. 
% jterasak@coma.ccs.tsukuba.ac.jp:
% /work/DBTHEORY/jterasak/schrpa/scnrpa/lipkin/nhrpans/nhrpans3.0/n=8/v=-0.5/rmix=0.1_converged/chi_selfconsistent
In fact, Eqs.~(\ref{eq:alphak0_3b}) and (\ref{eq:alphak0_3}) with the symmetry-conserving basis can be rewritten to an expression similar to Eq.~(\ref{eq:vacuum_condition_comp}), that is,  
\begin{align}
&\left(
\begin{array}{cccccc}
c_k                                  & L^{\prime k}_{12}         &                              & \cdots                    &           & L^{\prime k}_{1n} \\
L^{\prime k}_{21} & c_k                                 & L^{\prime k}_{23} & \cdots                    &           & L^{\prime k}_{2n} \\
L^{\prime k}_{31} & L^{\prime k}_{32} & c_k                        &                              &           & \vdots \\
\vdots                             &                                        &                             &                              &  \ddots & L^{\prime k}_{n-1\ n}\\
L^{\prime k}_{n1} & \cdots                              &                             &                              & L^{\prime k}_{n\ n-1} & c_k 
\end{array}
\right)
\left(
\begin{array}{c}
\beta_2/\beta_0 \\
\beta_4/\beta_0 \\
\vdots \\
\\
\beta_{2n}/\beta_0
\end{array}
\right) 
= 
\left(
\begin{array}{c}
-Y^k_2 \\
-Y^k_4 \\
\vdots \\
\\
-Y^k_{2n}
\end{array}
\right) , \label{eq:vacuum_condition_comp_even_rewritten}
\end{align}
\begin{align}
L^{\prime k}_{ij} &= X^k_{2j-2i} \sqrt{ \frac{G(2j)}{G(2j-2i)G(2i)} }, \ \ (i<j), \\
L^{\prime k}_{ij} &= Y^k_{2i-2j} \sqrt{ \frac{G(2i)}{G(2i-2j)G(2j)} }, \ \ (i>j).  
\end{align}
% Reformulation_vacuum_condition_even-order_Feb25_2017.pdf, Eqs.(4.4)-(4.6).

We investigated the truncation approximation with respect to $\tilde{J}_+$. 
%Figure \ref{fig:vw_ntr_N8} shows the results of the exact and the truncated nonlinear higher RPA calculations for $N=8$ in a region of $V$ having the symmetry-breaking HF ground state. Contrary to the expectation, the truncated approximation is not good at all. 
The correct solutions of the truncated nonlinear higher RPA were not obtained in the region of $V$ of the symmetry breaking in the HF approximation. 
The iteration process converged to an unphysical solution, e.g., the first excitation energy $E_{10}$ is 1.165 for $V=-0.2$ and the truncation order of 6 ($N=8$) when the shell-model truncated at the same order gives 0.703 (the exact value is 0.527). Unphysical solutions are possible because of the nonlinearity of the eigenequation. 
The reason for this difficulty can be discussed by analyzing the exact wavefunctions. Table \ref{tab:comp_exact} shows the squared norm of the lower-order components and that of the higher-order components for $V=-0.2$. The components in the symmetry-breaking basis are more distributed to the higher order than those in the symmetry-conserving basis. Therefore, the neglected components in the symmetry-breaking basis are more important than those in the symmetry-conserving basis. In addition, when the amplitudes have a broad distribution, it is difficult to have an input wavefunction close to the solution.  
The reason for that unexpected amplitude distribution can be inferred from the fact that the exact solution does not have the phase transition; in this model, the symmetry breaking is an artifact of approximation. Thus, one of the reasons for this problem is the property of the Lipkin model. 
%\begin{figure}[t]
%\centering
%\includegraphics[width=6.3cm]{vw_ntr_N8}
%\caption{ \label{fig:vw_ntr_N8} (Color online) $V$ dependence of the first excitation energy $E_{10}$ of the exact (solid line) and truncated nonlinear higher RPA calculation for $N=8$. For the exact calculation, $E_{10}=E^o_{10}$. The numbers in the figure indicate the truncation order with respect to $\tilde{J}_+$. }
%\end{figure}
% jterasak@coma.ccs.tsukuba.ac.jp:
% /work/DBTHEORY/jterasak/schrpa/scnrpa/lipkin/nhrpans/nhrpans3.0/n=8/v=-0.15/nhrpans3.2/chi_selfconsistent/v=-0.15_-0.30/ntr=4
\begin{table*}
\caption{\label{tab:comp_exact} Squared norm of components of lower and higher order with respect to $J_+$ (the symmetry-conserving basis) or $\tilde{J}_+$ (the symmetry-breaking basis) of the exact first-excited state with $V=-0.2$. The lower order in the symmetry conserving basis consists of the first and third (the higher is the fifth and seventh), and that in the symmetry-breaking basis is from the zeroth to the fourth (the higher is from the fifth to the eighth).}
%\begin{ruledtabular}
\centering
\begin{tabular}{ccc}
\hline
\multirow{2}{*}{Basis} &  \multicolumn{2}{c}{Squared norm of components} \\
                                  &  Lower order & Higher order                      \\
\hline
Symmetry conserving & 0.943  & 0.057  \\
Symmetry breaking    & 0.716  & 0.284 \\
\hline
\end{tabular}
%\end{ruledtabular}
\end{table*}
% jterasak@coma.ccs.tsukuba.ac.jp:
% /work/DBTHEORY/jterasak/schrpa/sc3rpa/lipkin/sc3rpa2x/sc3rpa2.1/n=8/voe=-0.150/partial_norm.s
% /work/DBTHEORY/jterasak/schrpa/scnrpa/lipkin/nhrpans/nhrpans3.0/n=8/v=-0.15/nhrpans3.2/chi_selfconsistent/v=-0.15_-0.30/ntr=6/exact_start/v=-0.20_exact_start/partial_norm.s
% /work/DBTHEORY/jterasak/schrpa/scnrpa/lipkin/nhrpans/nhrpans3.0/n=8/v=-0.15/nhrpans3.2/chi_selfconsistent/v=-0.15_-0.30/ntr=8/smlns2.0
% comp_cal.s

\section{\label{sec:summary}Summary}
We have shown that the nonlinear higher RPA reproduces the exact solutions of the Lipkin model.
The reconstruction has been demonstrated analytically for $N=2$ and numerically for $N=8$. 
Our study is the first one which shows the reproduction of the exact solutions of the Lipkin model for arbitrary $N$ within an extension of the RPA method. The proper construction of the phonon operator is crucial. 
We examined the results carefully and conclude that every property of the solutions is consistent. 
The only approximation in the realistic applications is the truncation of the wavefunction space, thus, 
there is no possibility of the lack of physical effects due to the mathematical properties of the nonlinear higher RPA.  
%It is also guaranteed, for example, that the second RPA approaches the exact method by using the ground state for calculating the matrices of the equation. 
%This prospect is applied to the calculations of any fermion many-body systems. 
Considering that the neck point of the shell model is its huge matrix dimension, the advantage of the nonlinear higher  RPA under the truncation is encouraging. 

We have also shown the formulation with the HF basis breaking the symmetry and reproduced the exact solution. 
However, the truncated solutions with this basis could not be obtained. It is an open question whether this problem occurs in the realistic systems having phase trnasitions. 
%It is expected that this founding will allow a new generation of the RPA
%calculations based on nonlinear phonon operators, which will  offer new insight into the nuclear
%structure problems. 

For the feasibility of the realistic calculation, there are calculations of nuclei by the second RPA \cite{Cat94,Pap10}. The feasibility of the iteration for solving the nonlinear second RPA is a matter of computational resource. Considering that the significant progress of the computers continues, the realistic application is a near-future task. It is also possible to consider a simplification \cite{Sme15} by introducing an approximate ansatz for the ground state of the nonlinear higher RPA.  

%\begin{acknowledgments}
\section*{Acknowledgments}
This work is supported by the VEGA Grant Agency of the Slovak Republic
under Contract No.~1/0922/16, the Ministry of Education, Youth and Sports
of the Czech Republic under Contract No.~LM2011027, RFBR Grant No.~16-02-01104, 
Underground laboratory LSM - Czech participation to European-level research infrastructure 
cz.02.1.01$/$0.0$/$0.0$/$16\_013$/$0001733, 
and Grant No.~HLP-2015-18 of the Heisenberg-Landau Program.
M.I.K.~acknowledges the support of Alexander von Humboldt Foundation.
This work is also supported by Bogoliubov Laboratory of Theoretical Physics, Joint Institute for Nuclear Research through the stay of one of the authors there. 
The computer COMA of Center for Computational Sciences, University of Tsukuba was used through the Interdisciplinary Public Program of fiscal year 2016 (TKBNDFT) for the numerical calculation. 
%\end{acknowledgments}

%\bibliography{nhrpa}
%\bibliography{scnrpa_lipkin}


\end{document}